\documentclass[hyper]{JHEP} 
\usepackage{epsfig}
\newcommand\fverb{\setbox\pippobox=\hbox\bgroup\verb}
\newcommand\fverbdo{\egroup\medskip\noindent%
            \fbox{\unhbox\pippobox}\ }

\newcommand\fverbit{\egroup\item[\fbox{\unhbox\pippobox}]}
\newbox\pippobox

\title{D-brane dynamics in a plane wave background}

\author{J. Kluso\v{n}\footnote{On leave from
Masaryk University, Brno, Czech Republic}\\
Dipartimento di Fisica,\\
Universita di Roma \& I.N.F.N. Sezione di ``Tor Vergata'' \\
Via della Ricerca Scientifica 1, 00133  Roma   ITALY\\

    E-mail: \email{Josef.Kluson@roma2.infn.it}}

\author{{Rashmi R. Nayak, and Kamal L. Panigrahi} \\
Dipartimento di Fisica (DIFI), \\
Universita di Genova \& I.N.F.N. Sezione di Genova \\
Via Dodecaneso 33, 16146 Genova, ITALY \\
Email: \email{rashmi.nayak@ge.infn.it, kamal.panigrahi@ge.infn.it}}

\preprint{GEF-TH-13/2005\\
ROM2F/2005/25\\
\hepth{0512159}}

\abstract{By using the Dirac-Born-Infeld action we study the
dynamics of D$p$-brane propagating in the NS5-near horizon plane
wave background. We study systematically D-brane embedding in this
pp-wave background, and analyze the equations of motion for
various auxiliary fields. We further discuss the motion of the
probe D$q$-brane in the presence of source D$p$-branes in this
plane wave background.}

\keywords{D-branes}

\def\bA{\mathbf{A}}
\def\bAi{\left(\mathbf{A}^{-1}\right)}
\def\mF{\mathcal{F}}
\begin{document}
\section{Introduction}\label{first}
Study of string theory in time dependent background is a
challenging topic which is believed to be able to answer questions
in early universe cosmology. One of the most studied subject in
this direction is the decaying phenomena of the unstable D-brane
in the presence of a tachyonic mode. The condensation of this
tachyon leads to a more stable brane configuration or a complete
annihilation (in case of brane-anti-brane pair). In the conformal
field theory language, it corresponds to studying the boundary
conformal field theory of the D-brane by a marginal deformation.
The recent proposal of Sen marks the spatially homogeneous decay
of the unstable D-brane by a deformation of the open string
worldsheet by an exact marginal rolling tachyon
background\cite{Sen:2002nu,Sen:2002in,Sen:2002an}. This process
can also be realized by the localization of S-brane in a time like
direction. In the study of the time dependent solutions in string
theory, the recent observations reveal that the Dirac-Born-Infeld
action captures, surprisingly well many aspects of the decay of
unstable D-branes\cite{Sen:1999md} -\cite{Kluson:2000iy}. More
recently a geometric tachyon has been prosed in
\cite{Kutasov:2004dj} that is the decay of the D$p$-brane into the
throat of a stack of NS5-branes. It has been observed that not
only the decay process resembles with that of the rolling tachyon
of the open string models, but also one could learn even more far
reaching consequences by studying in general the time dependent
dynamics of the D-branes in curved backgrounds. Hence one attempts
naturally to make more progress in the understanding of the
physics of D-branes in curved backgrounds.

In the recent past string theory in plane-wave background
\cite{Blau:2001ne} has also been a topic of intense discussion.
String theory in this background has been shown to be exactly
solvable in light-cone gauge and it provides a perfect laboratory
for testing the celebrated AdS/CFT duality beyond the supergravity
regime. String theory in pp-wave background that arises as the
Penrose limit of certain near horizon geometries is known to
provide a holographic description of certain sectors of dual field
theory \cite{Berenstein:2002jq}. Study of D-branes in this
background is interesting and has been investigated by using
various techniques in the past, see for
example:\cite{Billo:2002ff,Dabholkar:2002zc,
Kumar:2002ps,Skenderis:2002vf,Bain:2002nq,Bain:2002tq,
Biswas:2002yz,Nayak:2002ty,Sarkissian:2003jn}.
One of the interesting pp-background is obtained in the Penrose
limit of the near horizon region of a stack of NS5-brane (the
linear dilaton background) \cite{Hubeny:2002vf}. It is one of the
simplest pp-wave backgrounds with constant NS-NS flux and was
shown to resemble very closely to the flat space. D-brane
solutions in this background has been studied in
\cite{Hassan:2003ec}. While the perturbative spectrum seems to be
close enough to the flat space, the inclusion of non-perturbative
objects like D-branes changes drastically the situation. For
example, the spacetime supersymmetry seems to be lost in the
presence of D-branes in the pp-wave background of linear dilaton
geometry.

The rest of the paper is organized as follows. In section-2 we
give a very brief review of the near horizon geometry of the
NS5-branes and the description of D$p$-brane in its pp-wave
background. In section-3 we describe the DBI action of the
D$p$-brane in general background, and the nature of the equations
of motion. Then we study various embeddings of the D-branes in
pp-wave background. In particular, we discuss two types of branes,
namely the `longitudinal' branes (both $(u, v)$ directions along
the brane worldvolume) and the `transversal' branes (with one of
the lightcone direction $(u)$ along the brane). We suppose that
the worldvolume fields depend on $u$ only and derive solutions
for them. We further generalize
the situation by turning on appropriate gauge fields on the
worldvolume of the branes. In section-4, we study the relative motion of the
D$q$-brane in the pp-wave background in the presence of other
D$p$-brane source. We argue that how the embedding of branes
changes the relative motion of various D-branes. We
propse a particular kind of embedding where the brane motion
resembles with that of the flat space-time. In section-5 we
present our conclusions.

\section{General discussion of D$p$-branes in the near horizon limit of
NS5-branes and their pp wave limit} Non-gravitational theory such
as the Little string theory (LST) is an interesting examples of
non local theory. This theory arises on the world volume of the
NS5-brane when one takes the decoupling limit: $g_s\rightarrow 0$
with fixed $\alpha'$. To learn about the high energy spectrum of
this theory, which is not yet fully understood, one generally
advocates in terms of the dual field theory language, namely, in
terms of the string propagation in linear dilaton background. To
probe the high energy regime, the Penrose limit of the spacetime
has been very useful and intuitive. Let's recall some basic facts
about the linear dilaton background and its associated Penrose
limit.

The string frame metric, 3-form $H$ and dilaton of the
NS5-brane background are given by,
\begin{eqnarray}
ds^2 &=& -dt^2 + dy^2_5 + H(r)(dr^2 + r^2 (d\theta^2
+ \cos^2 \theta d\psi^2 + sin^2 \theta d\phi^2))\, \cr & \cr
H&=&N \epsilon_3\,,\>\>\> e^{2\phi} = g^2_s H(r)\, \>\>\> H(r) = 1+
{{Nl^2_s} \over {r^2}}\,,\nonumber
\end{eqnarray}
where $\epsilon_3$ is the volume form on the transverse $S_3$, $N$
is the NS5-brane charge and $H(r)$ is the harmonic function in the
transverse directions of the NS5-branes. The near-horizon geometry
corresponds to the the limit $r\rightarrow 0$ which removes the
$1$ in $H(r)$ and, on rescaling the time $(t=\sqrt{N}l_s \tilde
t)$, leads to the linear dilaton background,
\begin{eqnarray}
ds^2 = N l^2_s \big( -d\tilde{t}^2 + {{dr^2}\over {r^2}} +
\cos^2 \theta d\psi^2 + d\theta^2 + sin^2 \theta d\phi^2 \big)+
dy^2_5\,.\nonumber
\end{eqnarray}
The Penrose limit is then taken with respect to the null geodesic
along the equator ($\theta=0$) of the transverse $S^3$ resulting in
the following expressions for the metric and NS-NS 3-form
\cite{Hubeny:2002vf},
\begin{eqnarray}
\begin{array}{l}
ds^2 = 2du dv -\mu^2(z_1^2+z_2^2)(du)^2+\sum_{a=1}^8 dz^a dz^a\,,
\\[.2cm]
B_{12} = 2\mu u\,.
\end{array}
\label{NS5pp-bg}
\end{eqnarray}

To construct D$p$-brane in this particular background, one could
in principle follow various methods. The one that was adpoted in
\cite{Hassan:2003ec} was to write an ansatz for a particular D$p$-brane
solution in the NS5-brane near horizon pp-wave background and then to
solve the type IIB field equations of motion
along with the Bianchi identities. The supersymmetry variations
revealed the absence of any Killing spinors, and hence the solutions
are non-supersymmetric. But a careful analysis of the worldsheet
study revealed that the D-branes in NS5-near horizon pp-wave
background preserves as much supersymmetry as the flat space.
The contradiction was resolved by showing that all Fourier
modes of the allowed supersymmetry parameters depend on worldsheet
coordinate $\sigma$, while a local description in terms of the
space-time variable is blind to the extension of the string.

Recently in the study of open string tachyon condensation, the
Dirac-Born-Infeld action for D-branes have been very useful in
understanding the intriguing aspects of underlying physics. Hence
one wonders whether approaching the problem from this view point
help us in understanding the D$p$-branes more  in pp-wave
background. We would like to stress that the aim of this paper is
not to look for supersymmetric solutions in this plane wave
background, rather to study the properties of already known
D-brane solutions from a effective field theory point of view. We
consider various brane embedding in this pp-wave background and
study their dynamics.



Before going to the next section, where we discuss the DBI action
for a probe brane in NS5-pp wave background, we would like make a
bief review of the Penrose limit on the probe. The effect of the
Penrose limit on the dynamics of probe branes have been
investigated in \cite{Blau:2002mw}. It was observed that the
Penrose limit is essentially taking the large tension limit of the
probe brane. Below we give an out line following
\cite{Blau:2002mw} the analysis of the worldvolume dynamics of the
D$p$-brane in Penrose limit. Let us consider D$p$-brane in general
background
\begin{equation}
I_p[g,B,C]=-\tau_p\left[\int d^{p+1}\sigma
e^{-\Phi}\sqrt{-\det(g_{\mu\nu}+
\mF_{\mu\nu})}+\int e^{\mF}\wedge C\right] \ ,
\end{equation}
where
\begin{equation}
\tau_p=\frac{1}{(2\pi\alpha')^{\frac{p+1}{2}}k_p} \ ,
\end{equation}
where $k_p$ is a constant that
depends on string coupling constant
$g_s=e^{-\Phi(\infty)}$.
We als have
\begin{equation}
\mF_{\mu\nu}=\alpha'F_{\mu\nu}+B_{\mu\nu} \ ,
F_{\mu\nu}=\partial_\mu A_\nu-
\partial_\nu A_\mu \ ,
B_{\mu\nu}=B_{MN}\partial_\mu X^M
\partial_\nu X^N \ , C=\sum C_{(k)}
\end{equation}
Now we set $\alpha'=\Omega^2\alpha'$. The D$p$-brane action is
give by
\begin{eqnarray}
I_p \left[\Omega^{-2} g, \Omega^{-2} B, \Omega^{-k} C \right] &=&
\frac{1}{k_{p}{(2\pi\alpha')}^{\frac{p+1}{2}}}\Big(\int{d^{p+1}\sigma
e^{\Phi}\sqrt{\Omega^{-2}(g+B) + \alpha' F}}\cr
& \cr
&+& \left[\sum_{k} e^{\Omega^{-2}B + \alpha' F} \wedge \Omega^{-k} C_k\right]
\Big) \ .
\end{eqnarray}
Next step is to adopt the coordinates for the Penrose limit and take
$\Omega << 1$, the D$p$-brane  action is expanded in the following
way:
\begin{equation}
I_p\left[\Omega^{-2}g,\Omega^{-2}B, \Omega^{-k}C \right]
= I_p \left[\bar{g}, \bar{B},\bar{C}\right] + O(\Omega) \ ,
\end{equation}
where the bar valued quantities are the fields in the Penrose
limit. Hence one sees that the the D$p$-branes in the large
tension limit propagates in the Penrose limit of the associated
spacetime.

\section{D$p$-brane probe in NS5- near horizon pp-wave background}
Now we come to the main objective of the present paper, namely
we start to study the dynamics of probe D$p$-brane in the
NS5-near horizon plane wave background.
Recall that the action for a D$p$-brane in generic background
has the form
\begin{eqnarray}\label{actae}
S_p=-\tau_p\int d^{p+1}\sigma
e^{-\Phi}\sqrt{-\det \bA} \ ,
\bA_{\mu\nu}=\gamma_{\mu\nu}+F_{\mu\nu}
\ ,
\nonumber \\
\end{eqnarray}
where $\tau_p$ is the Dp-brane tension,
$\Phi(X)$ is dilaton, and
$\gamma_{\mu\nu} \ , \mu,\nu=0,\dots,
p$ is embedding of the metric to the worldvolume of D$p$-brane
\begin{equation}
\gamma_{\mu\nu}=g_{MN}\partial_\mu X^M
\partial_\nu X^N \ , M,N=0,\dots, 9
\end{equation}
In (\ref{actae}) the form $F_{\mu\nu}$ is defind as
\begin{equation}
F_{\mu\nu}=b_{MN}\partial_\mu X^M
\partial_\nu X^N+\partial_\mu A_\nu -
\partial_\nu A_\mu \ .
\end{equation}
The equation of motion for $X^K$ can be
easily determined from (\ref{actae})
and take the form
\begin{eqnarray}
\partial_K[e^{-\Phi}]\sqrt{-\det\bA}
+\frac{1}{2}e^{-\Phi}\left(
\partial_K g_{MN}
\partial_\mu X^M\partial_\nu X^N
+\partial_K b_{MN}\partial_\mu
X^M\partial_\nu X^N\right)\times \nonumber \\
\times \bAi^{\nu\mu}\sqrt{-\det\bA}
-\partial_\mu\left[e^{-\Phi}
g_{KM}\partial_\nu X^M\bAi^{\nu\mu}_S
\sqrt{-\det\bA}\right] - \nonumber \\
-\partial_{\mu}\left[e^{-\Phi}
b_{KM}\partial_\nu
X^M\bAi^{\nu\mu}_A\sqrt{-\det\bA}
\right]=0 \ ,
\nonumber \\
\end{eqnarray}
where, the symmetric and anti-symmetric part, respectively,
of the matrix $\bAi^{\mu\nu}$ is given by:
\begin{equation}
\bAi^{\nu\mu}_S=
\frac{1}{2}\left(\bAi^{\nu\mu}+
\bAi^{\mu\nu}\right)\ ,
\bAi^{\nu\mu}_A=
\frac{1}{2}\left(\bAi^{\nu\mu}-
\bAi^{\mu\nu}\right)\ .
\end{equation}
Finally, we should also determine the
equation of motion for the gauge field
$A_\mu$:
\begin{equation}\label{Aeq}
\partial_\nu \left[e^{-\Phi}\bAi^{\nu\mu}_A\sqrt{-\det\bA}
\right]=0 \ .
\end{equation}
Now we are going to study
the  time dependent dynamics of
the probe D$p$-brane in the plane wave
background of the linear dilaton
geometry that, as discussed in
the previous section, takes the form
\begin{equation}
ds^2= 2dudv-\mu^2(z_1^2+z_2^2)du^2 + \sum^{2}_{i=1}dz_i^2 + dx^2 +
dy_5^2 \ .
\end{equation}
together with nonzero NS-NS two form field
\begin{equation}
B_{12}=2 \mu u
\end{equation}
Recall that $u$ is the time coordinate. We will now solve the equations
of motion for D$p$-brane embedded in this background. It is clear that their
properties will strongly depend on the
embedding of these D$p$-brane in the pp-wave background. Since we
are interested in the time dependent
case, we fix the worldvolume time
coordinate $\sigma^0$ to be equal to
target space coordinate $u$. From the
structure of the background metric it
is natural to consider two cases. The
fist case corresponds to D$p$-brane that
wraps the $v$ direction as well and
we denote this D$p$-brane as $(u,v,p-1)$
that means that it is also extended in
$p-1$ spatial dimensions including some
dimensions from $y^p$. Since the metric
obtained above corresponds to the
massless geodesic moving at constant
$r$ it is natural to presume that the
probe D$p$-brane is located at some fixed
$x$. We also presume that $z^i$ coordinates are
transverse to the worldvolume of
D$p$-brane. The second case corresponds
to the probe D$p$-branes that are not extended in $v$ direction,
but the $u$ direction is along the worldvolume of the brane.
\subsection{$(u,v,p-1)$-branes}
In this case we propose the following gauge fixing:
\begin{equation}
u=\sigma^0\equiv t \ , v=\sigma^p \ ,
\sigma^{a}=y^a \ , a=1,\dots,p-1 \ .
\end{equation}
Consequently the matrix $\bA$ of eqn. (\ref{actae}) is given by:
\begin{equation}
\bA=\left(\begin{array}{ccc}
-\mu^2Z_iZ^i+\partial_0
Z_i\partial_0Z^i+(\partial_0
X)^2+\partial_0
Y^r\partial_0 Y_r & 0 & 1 \\
0  & \delta_{ab} & 0 \\
1 & 0 & 0  \\ \end{array}
\right)
\end{equation}
with $r, s = p, \dots, 5$ and $i=1, 2$.
Now looking at the form of the metric
and dilaton in the pp wave limit we see
that they are functions of $z^i$ only
and consequently the equations of
motion for $X$ and $Y^r$ take the forms
\begin{equation}
-\partial_0\left[e^{-\Phi}
g_{XX}\partial_0 X\bAi^{00}_S
\sqrt{-\det\bA}\right] = 0 \ ,
\end{equation}
Since $\bAi^{00}=0$ we see that
the equation of motion are obeyed
for any $X$. The same also holds
for $Y^r$ since
\begin{equation}
\partial_0\left[e^{-\Phi}
g_{rr}\partial_0 Y^r\bAi^{00}_S
\sqrt{-\det\bA}\right] = 0 \
\end{equation}
We should also solve the equation
of motion for $U,V$ and $Y^p$.
Since once again the metric does not
depend on these variables we get
\begin{eqnarray}
\partial_0\left[\sqrt{-\det\bA}\right]=0
\end{eqnarray}
that is again obeyed trivially. On the other hand the
equations of motion for $V$ and $Z^i$ are obeyed automatically
because of the vanishing  $\bAi^{00}$.

In summary, we see that for D$p$-brane of
the type $(u,v,p-1)$ the equations of
motion do not restrict possible form of
these D$p$-branes. Similar observations were made in
\cite{Skenderis:2002vf} for D$p$-branes
in maximally supersymmetric pp-wave background.

Before we proceed to the more general
ansatz we consider the possibility to have world volume modes
that depend on $v$ only. Consequently the matrix $\bA$ takes the
form
\begin{equation}
\bA=\left(\begin{array}{ccc}
-\mu^2Z_iZ^i & 0 & 1 \\
0  & \delta_{ab} & 0 \\
1 & 0 & g_{IJ}
\partial_p X^I\partial_p X^J  \\ \end{array}
\right)
\end{equation}
Then we get that the
determinant is equal to
\begin{equation}
\det\bA=-\mu^2Z_iZ^i(g_{IJ}
\partial_v X^I\partial_v X^J)-1 \ ,
\end{equation}
where we have used the gauge fixing $v=\sigma_p$.

As usual, the equations of motion
for $X$ and $Y^r$ take the form
\begin{eqnarray}
\partial_v\left[g_{xx}
\partial_v X\bAi^{vv}_S\sqrt{-\det\bA}
\right]=0 \ , \nonumber \\
\partial_v\left[g_{rs}
\partial_v Y^s\bAi^{vv}_S\sqrt{-\det\bA}
\right]=0 \ , \nonumber \\
\end{eqnarray}
that can be solved with the ansatz $\partial_vX=\partial_v Y^s=0$.
On the other hand the equation of motion for $Z^i$ is equal to
\begin{eqnarray}
\frac{\mu^2 Z_i(\partial_vZ^j
\partial_vZ_j)}{\sqrt{-\det\bA}}
-\partial_v\left[
\frac{\partial_v Z^i(\mu^2Z^jZ_j)}
{\sqrt{-\det\bA}}\right]=0
\nonumber \\
\end{eqnarray}
Let us propose the ansatz for $Z^1$ and $Z^2$ in the following form
\begin{equation}
Z^1=R\cos(kv) \ , Z^2=R\sin(kv)
\end{equation}
that implies
\begin{equation}
\det\bA= - 1 - \mu^2k^2R^4
\end{equation}
and the equation above takes the form
\begin{equation}
\frac{\mu^2Z_ik^2R^2}{
\sqrt{-\det\bA}}
-\partial_v^2Z\frac{\mu^2R^2}{\sqrt{-\det\bA}}=0
\end{equation}
that implies that $k=0$ and
consequently $Z^1=R \ , Z^2=0$.
On the other hand let us presume
that $Z^2=0$ (this solves the equation
of motion for $Z^2$ but for $Z^1=f(v)$.
Then $\det\bA$ is not a constant
and we should check that the equations
of motion for $U$ and $V$ are also obeyed.
The equation of motion for $U$ takes
the form
\begin{eqnarray}
\partial_v\left[(\mu^2 Z_1^2 - \mu^2 Z_1^2)
\frac{1}{\sqrt{-\det\bA}}\right]=0
\end{eqnarray}
and hence it is again obeyed.
On the other hand the equation of motion for $V$
takes the form
\begin{equation}
\partial_v \left[\frac{1}{\sqrt{-\det\bA}}\right]=0
\end{equation}
that implies
\begin{equation}
\sqrt{-\det\bA}= K \ ,
\end{equation}
where $K$ is a constant. Now from this equation we get the
differential equation for $Z^1$ in the form
\begin{equation}
Z_1dZ_1=\frac{\sqrt{K^2-1}}{\mu}dv
\end{equation}
that has the solution
\begin{equation}
\frac{Z_1^2}{2}+C=
\frac{\sqrt{ K^2- 1}}{\mu^2}v
\end{equation}
with some constant $C$. On the other hand the equation
of motion for $Z^1$ in case of constant $\det\bA$ takes the form
\begin{equation}
\mu^2Z_1(\partial_vZ^1)^2-\mu^2
\partial_v^2Z^1(\partial_vZ_1)^2
- 2 \mu^2(\partial_vZ^1)^2\partial_v^2
Z^1=0
\end{equation}
If we insert the ansatz given above
we get
\begin{equation}
\frac{\mu^2(K^2-1)}{Z}-
3\frac{\mu^2(K^2-1)^2}{Z^5}=0
\end{equation}
and we see that the equation of motion is obeyed for $K =1$ that
also implies $\partial_vZ=0$.

\subsection{More general ansatz}
In this section we generalize the solution given above to the case when
some modes depend on $u,v$ as well. Firstly we eliminate
some world volume fields where the metric explicitly does not depend on
them. More precisely, the equation of motion for $X$ takes the form
\begin{equation}
\partial_\mu\left[e^{-\Phi}g_{xx}\partial_\nu
X\bAi^{\nu\mu}\sqrt{-\det\bA}\right]=0
\end{equation}
that can be solved with $\partial_\mu
X=0$. At the same time we will solve
the equation of motion for $Y^r$ with
the same ansatz. In summary, $X$ and
$Y^r$ will be considered as constant.
In order to simplify calculation
further we will restrict in this paper
to the study of the D1 and D3-probe in
given background and we will study them
separately.
\subsection{$(u, v$)-D1 brane}
We will consider D1-brane that is extended in
$u,v$ directions. In this case the
non-zero components of the matrix $\bA$ takes the form
\begin{eqnarray}
\bA_{uu}= -\mu^2Z_iZ^i+(\partial_u
Z^i)^2 \ , \bA_{vv}=(\partial_v Z^i)^2, \nonumber \\
\bA_{uv}= 1 +\partial_uZ^i\partial_vZ_i
+\partial_uA_v+ 2 \mu u(\partial_u
Z^1\partial_vZ^2 -\partial_u
Z^2\partial_v Z^1) \ , \nonumber \\
\bA_{vu}= 1 +\partial_vZ^i\partial_uZ_i
-\partial_uA_v+2 \mu u(\partial_v
Z^1\partial_uZ^2 -\partial_v
Z^2\partial_u Z^1)  \ ,
 \nonumber
\\
\end{eqnarray}
where we presume that all free fields on the
world volume of D1-brane are constant.

We start to solve the equation of motion for $U$ and $V$.
The equation of motion for $U$ takes the form
\begin{eqnarray}\label{eqUga}
\partial_u\left[(\mu^2Z_iZ^i\bAi^{uu}_S+2\bAi^{vu}_S)
\sqrt{-\det\bA}\right]- \nonumber \\
\partial_v\left[(\mu^2Z_iZ^i\bAi^{uv}_S+2
\bAi^{vv}_S)\sqrt{-\det\bA}\right]=0 \ .
\end{eqnarray}
Similarly the equation of motion for $V$ takes the form
\begin{equation}\label{eqVga}
\partial_u\left[
\bAi^{uu}_S\sqrt{-\det\bA}\right] +
\partial_v\left[
\bAi^{uv}_S\sqrt{-\det\bA}\right]=0 \ .
\end{equation}
then we get following equation of
motion for $Z^1$
\begin{eqnarray}
-\mu^2Z^1\bAi^{uu}\sqrt{-\det\bA}
-\partial_\mu\left[\partial_\nu
Z^1\bAi^{\nu\mu}_S\sqrt{-\det\bA}\right] \nonumber \\
-\partial_\mu\left[b_{12}\partial_\nu Z^2
\bAi^{\nu\mu}_A\sqrt{-\det\bA}\right]=0
\end{eqnarray}
and clearly the same  for $Z^2$.
Let us now propose the ansatz for $Z^1$ and $Z^2$ as follows
\begin{equation}
Z_1=R\cos(k(u+v)) \ , Z_2=R\sin(k(u+v))
\ .
\end{equation}
Looking at the form of the equation
given above we see that in case
of nonzero $A_v$ they take very complicated
form thanks to the explicit time dependence of $b_{12}$. For
that reason we restrict to the case
when $\partial_\mu A_v=0$. Then we get that $\bAi_A=0$. Explicitly
\begin{equation}
\bA=\left(\begin{array}{cc}
-\mu^2R^2+k^2R^2 & 1+k^2R^2 \\
1+k^2R^2 & k^2R^2 \\
\end{array}\right)
\end{equation}
and hence
\begin{equation}
\det\bA=-1 - R^2k^2(2+\mu^2 R^2)
\end{equation}
Since now $\bAi \ , \det\bA \ , Z_iZ^i$ are constant it is easy to
see that the equations of  motion (\ref{eqUga}) and (\ref{eqVga})
are trivially satisfied. On the other hand the equation of motion
for $Z^i$ takes the form (using the fact that $\partial_\mu Z^1=
-kR\sin(k(u+v))$
\begin{eqnarray}
\frac{Z_1}{\sqrt{-\det\bA}}
\left(\mu^2k^2R^2-4k^2+\mu^2R^2k^2\right)=0
\end{eqnarray}
This however implies that $k=0$ and we get $Z^1=R={\rm const}$ and $Z^2=0$.

As the next possibility we will
consider the following ansatz for the $Z^1$ and $Z^2$
\begin{equation}
Z^1=R\cos(k(u-v)) \ , Z^2=R\sin(k(u-v)) \
\end{equation}
that implies following form of
the matrix $\bA$
\begin{equation}
\bA=\left(\begin{array}{cc}
-\mu^2R^2+R^2k^2 & 1-k^2R^2 \\
1-k^2R^2 & k^2R^2 \\ \end{array}
\right)
\end{equation}
and hence
\begin{equation}
\det\bA=-1-k^2R^2(\mu^2 R^2 -2) \ .
\end{equation}
Again, the nontrivial
equation of motion corresponds to $Z^{1,2}$and takes the form
(using the fact that $\partial_u Z^1= -kR\sin(k(u-v)) \ ,
\partial_v Z^1=kR\sin(k(u-v))$)
\begin{eqnarray}
\frac{2Zk^2}{\sqrt{-\det\bA}} k^2(1-\mu^2 R^2)=0 \ .
\end{eqnarray}
Now we see that the above equation has two solutions.
One with $k=0$, and other with arbitrary $k$ but with $R=\mu^{-1}$.

\subsection{$(u, v ,2)$-D3 brane}
We now consider D3-brane where additional spatial components span $y$
subspace. More precisely, we define the embedding of this D3-brane as
\begin{equation}
U=\sigma^0=u \ , V=\sigma^3=v \ ,
y^1=\sigma^1 \ , y^2=\sigma^2 \ .
\end{equation}
Then the embedding coordinates are $X,
Y^r \ , r=3,4,5$ and $Z^i, i=1,2$. We
also switch on the gauge field
$A_{2,3}$ where, following results
given in previous subsection, we take
$A_v=0$. Using $SO(2)$ symmetry in the
subspace spanned by $y^1,y^2$ we will
switch on the component $A_1$ only and
we will presume that this field depends
on $u$ only. Again the  equations of motion
for $X,Y^r$ will be solved with the
ansatz $X,Y^r$ are constant.
As in the previous subsection we take the ansatz for $Z^i$ modes as
\begin{equation}
Z^1=R\cos(k(u+v)) \ , Z^2=R\sin(k(u+v)) \
\end{equation}
We again get
\begin{equation}
\bA=\left(\begin{array}{cccc}
-\mu^2R^2+k^2R^2 & \partial_uA_1 & 0 & 1 + k^2R^2 \\
-\partial_uA_1 & 1 & 0 & 0 \\
0 & 0 & 1 & 0 \\
1+k^2R^2 & 0 & 0 & k^2R^2 \\
\end{array}\right)
\end{equation}
Then the determinant takes the form
\begin{eqnarray}
\det\bA = -\mu^2k^2R^4 - 1 - 2k^2R^2+(\partial_uA_1)^2
k^2R^2 ,
\end{eqnarray}
Firstly we start to solve the equation
of motion for $A_\mu$. These equations
take the form
\begin{equation}
\partial_\mu\left[\bAi^{\nu\mu}_A
\sqrt{-\det\bA}\right]=0
\end{equation}
We see that all equations of motion
are solved with where $\bA$ is constant.
This implies that
\begin{equation}
\partial_u A_1\equiv n=const \ .
\end{equation}
Generally, $n$ could take any real value
however it is well known that it is proportional to the number of
fundamental strings.

Now we have that the matrix $\bA$ and its
inverse is constant. This implies
that we could proceed as in the
previous section. However we should
check that the term containing the
target space $b_{12}$ in the equations
of motion vanishes. In fact, this
term is
\begin{equation}
\partial_{\mu}\left[
b_{12}\partial_\nu \bAi^{\nu\mu}_A
\sqrt{-\det\bA}\right]
\end{equation}
that for $Z$ that depends on $u,v$ gives
\begin{eqnarray}
\partial_u\left[b_{12}
(\partial_u Z^2\bAi^{uu}_A+
\partial_vZ^2\bAi^{vu}_A)\sqrt{-\det\bA}
\right]+\nonumber \\
\partial_v\left[
b_{12}(\partial_u Z^2
\bAi^{uv}_A+\partial_vZ\bAi^{vv}_A)
\sqrt{-\det\bA}\right]=0
\nonumber \\
\end{eqnarray}
using the fact that $\bAi^{uu}_A=\bAi^{vv}_A
=\bAi^{uv}=0$. Then we can really
proceed as in the previous
subsection and the equation of motion
for $Z^1$ gives
\begin{eqnarray}
\frac{Z_1}{\sqrt{-\det\bA}}
\left(- 2 + n^2 + \mu^2 R^2\right) k^2=0 \ .
\nonumber \\
\end{eqnarray}
so that we once again obtain the condition $k=0$.
\subsection{$(u,\emptyset,p)$ brane}
Now let us discuss the case, when the D$p$-branes don't wrap the
$v$-direction. We call them $(u,\emptyset,p)$ branes. In this case,
the static gauge has the form
\begin{equation}
U=\sigma^0\equiv u,\sigma^a=y^a \ , a=1,\dots,p
\end{equation}
and hence the embedding coordinates are
\begin{equation}
V,X,Z^i,Y^r \ , r=p+1,\dots, 5 \ .
\end{equation}
In this subsection we will consider
more general case when $p=1$ and $p=3$.
Then the non-zero components of the matrix $\bA$ are
\begin{eqnarray}
\bA_{uu}= -\mu^2Z_iZ^i +2 \partial_u V
+(\partial_u X)^2+(\partial_u Z^i)^2+
(\partial_u Y^i)\ , \bA_{ab}=\delta_{ab} \ . \nonumber \\
\end{eqnarray}
Consequently we get
\begin{equation}
\det \bA=-\mu^2Z_iZ^i+2\partial_u V
+(\partial_u X)^2+(\partial_u Z^i)^2+
(\partial_u Y^i)^2
\end{equation}
Again we start to solve the equation of motion. For $U=\sigma^0$
we get
\begin{eqnarray}\label{Usigma}
\partial_u\left[
\frac{\left(-\mu^2Z_iZ^i+\partial_u V\right)}{
\sqrt{ \mu^2Z_iZ^i-2\partial_u V
-(\partial_u X)^2-\partial_u Z^i
\partial_u Z_i -
\partial_u Y^i\partial_u Y_i}}
\right]=0 \ . \nonumber \\
\end{eqnarray}
The equations of motion for $Y^a=\sigma^a$ imply
\begin{equation}
\partial_u\left[g_{ab}
\bAi^{bu}\sqrt{-\det\bA}\right]=0
\end{equation}
using the fact that all modes depend on $t$ only.
The equation of motion for $X$ imply
\begin{equation}
\partial_u\left[\frac{\partial_u X}
{\sqrt{ \mu^2Z_iZ^i-2\partial_u V
-(\partial_u X)^2-\partial_u Z^i
\partial_u Z_i -
\partial_u Y^i\partial_u Y_i}}\right]=0
\end{equation}
which in turn specify the conserved momentum $P_x$:
\begin{equation}
\frac{\partial_u X}
{\sqrt{ \mu^2Z_iZ^i - 2\partial_u V
-(\partial_u X)^2-\partial_u Z^i
\partial_u Z_i -
\partial_u Y^i\partial_u Y_i}}=P_x \ .
\end{equation}
In the same way we  obtain conserved momenta $P_r$
\begin{equation}
\frac{\partial_u Y^r}
{\sqrt{ \mu^2Z_iZ^i- 2 \partial_u V
-(\partial_u X)^2-\partial_u Z^i
\partial_u Z_i -
\partial_u Y^i\partial_u Y_i}}=P_r
\end{equation}
Finally, the equation of motion
for $V$ is equal to
\begin{equation}
\partial_u\left[
\frac{1}{\sqrt{-\det\bA}}\right]=0 \ .
\end{equation}
In other words we get that
$\det\bA={\rm const}$. This also implies
$\partial_u V=const$ and hence the
equation (\ref{Usigma}) again
implies $Z^iZ_i = {\rm const}$.
Finally, the equation of motion
for $Z_i$ is equal to
\begin{eqnarray}
\partial_u^2Z^i + \mu^2 Z^2_i=0
\end{eqnarray}
that has again the solution
\begin{equation}\label{Zi}
Z^1=R_0\sin \mu u \ ,Z^2=R_0\cos\mu u \ .
\end{equation}
In  summary, we obtain following dependence of the worldvolume
fields
\begin{equation}
X=V_x u+X_0 \ ,
V=V_v u+V_0 \ ,
Y^r=V_{r}u+Y^r_0 \ .
\end{equation}
together with the time dependence of $Z^i$ given in (\ref{Zi}).
\section{D-brane motion}
In this section, we would like to analyze the
motion of probe D$q$-brane in the presence of another (or a stack of)
D$p$-brane in this particular pp-wave background.
We assume that the dimension of the background brane
is always greater than that of the probe. This
is a good approximation when the mass of the
higher dimensional brane is bigger than the
mass of the lower dimensional brane. To perform
the analysis we start by writing down the classical solution of a
D$p$-brane in pp-wave background, and try to probe a D$q$-brane.
The metric, dilaton and the NS-NS 3-form flux of such a
configuration is given by \cite{Hassan:2003ec}
\begin{eqnarray}
&& ds^2=H^{-{1\over 2}}_p
(x^a)\left[2dudv -
\mu^2\sum^2_{i=1}z^2_i du^2 +
\sum^{p-1}_{\alpha=1}(dx^{\alpha})^2\right]
+H^{{1\over 2}}_p (x^a)\left[\sum^{2}_{i=1}(dz^i)^2+
\sum^{8}_{a=p+3}{(dx^a)}^2\right], \cr & \cr
&& B_{12} = 2\mu u \,, \>\>\> A^{(p+1)}
= \pm \left(\frac{1}{H(x^a)}-1\right)
du\wedge dv\wedge dx^1 \wedge
dx^2\wedge \dots \wedge dx^{p-1} \cr &
\cr && e^{2\phi} =
H_p(x^a)^{\frac{3-p}{2}}\,, \>\>\> H_p
= 1 + N g_s
\left(\frac{l_s}{r}\right)^{7-p}\equiv
1+\frac{\lambda_p}{r^{7-p}} \,.
\label{Dp}
\end{eqnarray}
where
\begin{equation}
r^2=\sum_{i=1}^2
z_iz^i+\sum_{a=p+3}^8x_ax^a
\end{equation}
The harmonic function written in
the last line for $p=7$, is given by
\begin{equation}
H_7 = 1-Ng_s \log(r/l).
\end{equation}
We will study the motion of the probe D$q$-brane, where
$q<p, p-q=2k, k=0,1,2$ in this
background. For simplicity we
consider the case when all the gauge fields
are set to zero. In general the action for such a D$q$-brane probe
is given by
\begin{equation}\label{dbi-wz}
S_=-\tau_q\int d^{q+1}\sigma
e^{-\Phi}\sqrt{-\det\bA}+S_{WZ}
\end{equation}
where
\begin{equation}
\bA_{\mu\nu}=g_{\mu\nu}+F_{\mu\nu} \ ,
\gamma_{\mu\nu}=g_{MN}\partial_\mu X^M
\partial_\nu X^N \ , M,N=0,\dots, 9 \ ,
F_{\mu\nu}=b_{MN}\partial_\mu X^M
\partial_\nu X^N .
\end{equation}
We propose the static gauge in the form
\begin{equation}
\sigma^0=u \ , \sigma^q=v \ ,
\sigma^i=x^i \ , i=1,\dots,q-1
\end{equation}
and try to see whether they solve the
appropriate equations of motion.
Generally we presume that the world volume fields depend on $u,v$
only. Then the non-zero components of the matrix $\bA$ are
\begin{eqnarray}
\bA_{00}= -H_p^{-1/2}\left(\mu^2z_iz^i+ \partial_0 Y^r\partial_0 Y_r\right)
+H^{1/2}\left(\partial_0Z_i\partial_0Z^i
+ \partial_0 X^a\partial_0 X_a\right) \nonumber \\
\bA_{0q}
= H^{-1/2}+H^{-1/2}
\partial_0 Y^r\partial_q
Y_r+H^{1/2}(\partial_0Z^i\partial_qZ_i
+\partial_0X^a\partial_q X_a)\nonumber \\
+\mu U(\partial_0 Z^1\partial_qZ^2-
\partial_0Z^2\partial_qZ^2)
 \nonumber \\
\bA_{q0}=
H^{-1/2}+H^{-1/2}
\partial_0 Y^r\partial_q
Y_r+H^{1/2}(\partial_qZ^i\partial_0Z_i
+\partial_0X^a\partial_q X_a)\nonumber \\
-\mu U(\partial_0 Z^1\partial_qZ^2-
\partial_0Z^2\partial_qZ^2) \ ,
\nonumber \\
\bA_{ij}=g_{ij}= H^{-1/2}\delta_{ij} \ ,
\nonumber \\
\bA_{qq}=
H^{-1/2}\partial_q Y^r\partial_q
Y_r+H^{1/2}(\partial_qZ^i\partial_qZ_i
+\partial_qX^a\partial_q X_a) \nonumber \\
\end{eqnarray}
where $r,s=q,\dots,p-1$. Since the metric does not depend on
$Y^r$ then the equations of motion for them take the form
\begin{equation}
\partial_\mu\left[e^{-\Phi}
g_{rs}\partial_\nu Y^s \bAi^{\nu\mu}_S
\sqrt{-\det\bA}\right]=0
\end{equation}
that can be solved with $Y^r={\rm const}$.
The problem seems rather complicated
thanks to the dependence of the metric on $x,z$. Since we are
interested in the properties of D$q$-brane as a probe it is
natural to restrict to the dependence of all modes on $\sigma^0$
only. In this case the non-zero components of the matrix $\bA$ are
\begin{eqnarray}
\bA_{00}=
-H_p^{-1/2}\left(\mu^2z_iz^i+\partial_0 Y^r\partial_0
Y_r\right)+H^{1/2}\left(\partial_0Z_i\partial_0Z^i+
\partial_0 X^a\partial_qX_a\right) \ , \nonumber \\
\bA_{0q}= H^{-1/2} = \bA_{q0} \ ,
\bA_{ij}=
H^{-1/2}\delta_{ij} \ ,
\bA_{qq}=0 \nonumber \\
\end{eqnarray}
this however implies that
$\det \bA=-H_p^{-\frac{q+1}{2}}$
Let us start to solve the equation of motion for $U$ that
is give by
\begin{eqnarray}
\partial_0\left[H_p^{\frac{p-q-4}{4}}
\right]=0 \ .  \end{eqnarray}
This result is trivially satisfied for $p-q=4$ while for $p-q=2$ this
is obeyed for $\partial_0 X^a= \partial_0 Z^i=0$.

As the next step we consider the equation of motion for $V$ that for
our ansatz takes the form
\begin{equation}
\partial_0\left[e^{-\Phi}g_{VU}
\bAi^{00}\sqrt{-\det\bA}\right]=0
\end{equation}
that are satisfied since $\bAi^{00}=0$.
Finally, the equation of motion for
$X^a$ take the form
\begin{eqnarray}
\partial_{X^a} e^{-\Phi}\sqrt{-\det\bA}+
\frac{1}{2}\partial_{X^a} g_{MN}
\partial_{\mu}X^M\partial_\nu X^N
\bAi^{\nu\mu}\sqrt{-\det\bA}=
\nonumber \\
=\lambda_p
(p-7)\frac{X^a}{R^{9-p}}H_p^{\frac{p-q-8}{4}}
\left(p+q-2\right)=0 \nonumber \\
\end{eqnarray}
using the fact that $\bAi^{00}=0$ and also
\begin{eqnarray}
\partial_{X^a}g_{MN}\partial_\mu
X^M\partial_\nu X^N\bAi^{\nu\mu}
= \frac{1}{2H^{3/2}}\frac{\delta H_p} {\delta
X^a}(q+1)\frac{H_p^{-\frac{q}{2}}}{\sqrt{-\det\bA}} \ ,
\nonumber \\
\partial_{X^a}e^{-\Phi}=
\frac{p-3}{4}H_p^{\frac{(p-7)}{4}}\frac{\delta
H_p}{\delta X^a} \ , \nonumber \\
\end{eqnarray}
It is clear that the same equation of
motion holds for $Z^i$ as well thanks
to the fact that the variation
$\frac{\delta g_{UU}}{\delta Z^i}$ is
proportional to $\bAi^{00}$ in the
equation of motion and this vanishes.
Let us then concentrate on the equation
above and discuss its properties for various
values of $p$ and $q$ and for limits $X^a\rightarrow 0$ or
$X^a\rightarrow \infty$. Generally, for
 $X^a\rightarrow 0$ that (for $Z^i=0, X^b=0 \ , b\neq a$) we have
\begin{equation}\label{lim}
\lim_{X^a\rightarrow 0}
\frac{X^a}{R^{9-p}}H_p^{\frac{p-q-8}{4}}
\sim \lim_{X^a\rightarrow 0}(X^a)^{\frac{1}{4}\left[(p-8)(p-3)-q(p-7)\right]}
\end{equation}
On the other hand the limit
$X^a\rightarrow \infty$ gives
\begin{equation}\label{limin}
\lim_{X^a\rightarrow \infty}
\frac{X^a}{R^{9-p}}H_p^{\frac{p-q-8}{4}}
\sim \lim_{X^a\rightarrow \infty}
\frac{1}{(X^a)^{9-p}}
\end{equation}
that goes to zero for all $p<7$. Let's discuss the situations case by case.
\begin{itemize}
\item {\bf $p=6$} \\
In this case $q=2,4$ since by presumption D$q$-brane wraps $u,v$
directions and, hence has to be at least two dimensional. For
$q=2,4$ we get that (\ref{lim}) blows up so that the point $X^a=0$
cannot be solution of the equation of motion. Then the only
possibility is to consider the configuration where all
$X^a\rightarrow \infty$. In other words D$q$-brane cannot form a
bound state with D6-brane. This result can be compared with the
analysis performed in \cite{Burgess:2003mm} where it was also
shown that in the near horizon region of D6-brane the potential
diverges.
\item {\bf $p=5$} \\
Now  $q=3,1$. Then the exponent on
$X^a$ is equal to $\frac{-3+q}{2}$ and hence we
again get the the expression
(\ref{lim}) diverges for $q=1$ while it
is constant for $q=3$. In any case
D3-brane or D1-brane cannot approach
D5-brane in this particular
configuration.
\item {\bf $p=4$}\\
Now in the limit $X^a\rightarrow 0$, we
have $(X^a)^{\frac{-4+3q}{4}}$, which vanishes
for $q=2$. Hence D2-brane can approach to D4-brane.
\item {\bf $p=3$}\\
In the limit $X^a\rightarrow 0$, one has ${(X^a)}^q$, which
for $q=1$ goes to zero and hence the D1-brane can approach
the D3-brane.
\end{itemize}

As the final example we will study the
probe Dp-brane in the background of
Dp-brane in the NS5-brane pp wave. In
this case we should take the WZ term
into account that takes the form
\begin{equation}
S_{WZ}=-\tau_p\int A^{(p+1)}=
-q_p\tau_1 \int d^{p+1}\sigma \left(
\frac{1}{H_p}-1\right)
   \ ,
\end{equation}
where $q_p$ takes values $\pm 1$
according to the case whether Dp-brane
probe corresponds to Dp-brane or anti
Dp-brane. The presence of this WZ term
only changes the equations of motion
for $X^a$ and $Z^i$ and we get
\begin{eqnarray}
\lambda_p (p-7)\frac{X^a}{R^{9-p}H_p^2}
\left(2p-2\right)-\lambda_p(p-7)\frac{q_pX^a}{R^{9-p}
H_p^2}= \lambda_p
(p-7)\frac{X^a}{R^{9-p}H_p^2}
\left(2p-2-q_p\right)=0 \nonumber \\
\end{eqnarray}
Then for $X^a\rightarrow 0$ the upper
expression is proportional to
$(X^a)^{6-p}$ that goes to zero for
$p<6$ and hence the only solution of
the equation of motion corresponds to
$X^a=Z^i=0$.


\subsection{Alternative embedding}
In this section we mention a possibility of an alternative
embedding of the probe D$q$-brane and examine the motion
of the probe brane described in the previous section.
Once again, we start with the action (\ref{dbi-wz}). However,
we propose the static gauge in the following form
\begin{equation}
U=\sigma^0+\sigma^q \ ,
V=\sigma^0-\sigma^q  \ , \sigma^i=X^i \
, i=1,\dots,q-1
\end{equation}
and try to see whether they solve the
appropriate equations of motion.We will also presume that the
world volume fields depend on $\sigma^0$ only. Then the matrix $\bA$
takes the form
\begin{eqnarray}
\bA_{00}
=-H_p^{-1/2}\left(\mu^2Z_iZ^i+ 2 + \partial_0Y^r\partial_0Y_r\right)
+H_2^{1/2}((\partial_0Z^i)^2+(\partial_0
X^a)^2) \ , \nonumber \\
\bA_{0q} = \bA_{q0} = -H_p^{-1/2}\mu^2Z_iZ^i
\ , \bA_{ij}=H^{-1/2}\delta_{ij}
\ ,\nonumber \\
\bA_{qq}= - H_p^{-1/2}\mu^2Z_iZ^i-2H_p^{-1/2}  \ .
\nonumber \\
\end{eqnarray}
so that the determinant is equal to
\begin{eqnarray}
\det\bA
= -H_p^{-(q+1)/2}\left[4+(\mu^2Z_iZ^i+2)((\partial_0Y^r)^2 +
H_p(\partial_0X^a)^2+
H_p(\partial_0 Z^i)^2)\right] \ , \nonumber \\
\end{eqnarray}
We again start with the equation of
motion for $U$ that now takes the form
\begin{eqnarray}
-4\partial_0\left[\frac{H_p^{(p-q-4)/4}(\mu^2Z_iZ^i+2)}
{\sqrt{4+(\mu^2Z_iZ^i+2)
((\partial_0Y^r)^2 +
H_p(\partial_0X^a)^2+ H_p(\partial_0
Z^i)^2)}}\right]=0
\nonumber \\
\end{eqnarray}
This equation has the form of the energy conservation equation
and in fact it can be interpreted as a conservation of
the world volume energy\cite{Aganagic:1996nn}.
We will use this equation later.
The equation of motion for $V$ takes the form
\begin{eqnarray}
\partial_0\left[
\frac{H_p^{(p-q-4)/4}}
{\sqrt{4+(\mu^2Z_iZ^i+2)
((\partial_0Y^r)^2 +
H_p(\partial_0X^a)^2+ H_p(\partial_0
Z^i)^2)
}}\right]=0 \nonumber \\
\end{eqnarray}
Now comparing the above two equations
we get the condition
\begin{equation}
\partial_0[\mu^2Z_iZ^i+2]=0 \ .
\end{equation}
To find the solution of the above differential
equation, let us consider the equation of motion for
$Z^i$. In this case the situation
is more complicated since we
have dilaton and metric component
$Z^i$ dependent. However since
the variation of $g_{uu}$ with
respect to $Z^i$ contains a
linear term $Z^i$ we will solve
the equation of motion for $Z^i$
in terms of $Z^i=0$ that is clearly
solution of the equation of motion.

Let us now consider the equation
of motion for $Y^r$ that takes the form
\begin{equation}
\partial_0\left[
e^{-\Phi}g_{rs}\partial_0Y^r
\bAi^{00}\sqrt{-\det\bA}\right]=0
\end{equation}
The solution of this equation of motion
is given by constant expression under
the right bracket that corresponds to
the constant momenta conjugate to
$Y^r$. If we could in principle express
$\partial_0Y^r$ with this momenta
and insert it to square root we
will consider the simpler case
when $P_r=0$ in order to not to
complicate the expressions further.

Now we come to the analysis of the dynamics of the modes $X^a$.
Using the manifest $SO(9-p)$ invariance of the
space $R^{9-p}$ transverse to the background Dp-brane we can restrict to
study of the dynamics in the two
dimensional plane, say $x^7, x^8$ where
we introduce coordinates
\begin{equation}
X^7=R\cos \theta \ ,
X^8=R\sin \theta \ .
\end{equation}
Now we could again in principle
solve the equation of motion for
$R,\theta$ directly, however we rather
use the equation of motion for $U$ that
has the form of the equation of the
conservation of the energy. In other
word, this equation implies
\begin{equation}
\frac{H_p^{(p-q-4)/4}} {\sqrt{4+2
(H_p(\dot{R}^2+R^2
\dot{\theta}^2)}}=\frac{E}{\tau_q} \ ,
\end{equation}
where $\dot(\dots)=\partial_0(\dots)$.
Again, since the action does not depend explicitly
on $\theta$ it turns out that the momentum conjugate
to $P_\theta$ is a constant. For simplicity we restrict to the
case of $P_\theta=0$. Now the equation above implies
\begin{equation}\label{eqc}
\frac{\dot{R}^2}{2}+
H_p^{-1}-\frac{\tau_q^2}{4E^2}
H_p^{\frac{p-q-4}{2}-1}=0
\end{equation}
that corresponds to the motion of the
particle with zero energy in the
potential of the form
\begin{equation}
V(R,E)=H_p^{-1}-\frac{\tau_q^2}{4E^2}
H_p^{\frac{p-q-4}{2}-1} \ .
\end{equation}

Now let us try to solve the above equation. However, we
can see that solving the equation above in full generality
is very hard. What we can do of course is to take the following
two simple possibilities. $q=p-2\Rightarrow p-q=2$ or $q=p-4
\Rightarrow p-q=4$ (the BPS case). In the second case the
differential equation above takes very simple form
\begin{equation}\label{pq4}
dRR^{(p-7)/2} =\pm\left(
\sqrt{\frac{\tau_q^2}{2E^2}-2}
\sqrt{\lambda_p}\right)d\sigma^0 \ ,
\end{equation}
where we have taken  the near horizon approximation where
$\frac{\lambda_p}{R^{7-p}}\gg 1$. Now the equation (\ref{pq4})
can be easily solved with the following (for $p\neq 5$)
\begin{equation}
R^{\frac{p-5}{2}}= \frac{p-5}{2}
\left(\left(\sqrt{\lambda_p}
\sqrt{\frac{\tau_q^2}{2E^2}-2}\right)
\sigma^0+C_0 \right)
\end{equation}
>From this result we see that for $p=6$, the  D2-brane
cannot reach the worldvolume of D6-brane in the same way as
in flat space time. For $p=5$, however, the equation above has the solution
\begin{equation}
R=R_0\exp \left(- \left(\sqrt{\lambda_p}
\sqrt{\frac{\tau_q^2}{2E^2}-2}\right)
\sigma^0\right) \ ,
\end{equation}
where we have chosen the $-$ sign in the exponential function to
find solution that describes D1-brane that approaches D5-brane.
Finally, we cannot consider $p<5$ in this case since we then have
$q<1$ however we have presumed that D$q$-brane is two dimensional
at least.

Let us now consider the first case when
$p-q=2$. Then the equation
(\ref{eqc}) implies following bound on
$R$
\begin{equation}
R^{7-p}>\frac{\lambda_p}{\frac{\tau_q^2}{4E^2}-1}
 \ .
 \end{equation}
In other words the D(p-2)-brane that
approaches Dp-brane from $r=\infty$ can reach the minimal distance
$R_{min}^{7-p}= \frac{\lambda_p}{\frac{\tau_q^2}{4E^2}-1}$.
In other words D$q$-brane that moves only radially cannot
reach the world volume of D$p$-brane.
\section{Conclusion}
In this paper we have discussed the dynamics of the D$p$-branes in
the NS5-near horizon pp-wave background. We consider various
embedding of the D-branes in this background to understand the
relevant physics out of it. We consider the longitudinal and the
transversal branes and study their dynamics. We observe that for
the branes which wrap both $u, v$ directions, the equations of
motion derived from the DBI action does not restrict the possible
form of the D-branes. We further analyze the properties of the
solutions by turning on additional worldvomune gauge fileds.
Finally we study the motion of probe D$q$-brane in the presence
of source D$p$-branes in this plane wave background. By assuming
that the background branes are heavier than the probe, we solve
the equations of motion derived from the DBI action of such probe.
Once again we have begin with the assumption that the worldvolume
fields depend on both $u$ and $v$. We explain that how the gauge
fixing plays a crucial role in the study of the brane motion.
By taking various examples, we
have also shown the interesting trajectory of the probe branes
falling into source branes in the pp-wave background of the linear
dilaton geometry. In a particular gauge fixing, we solve the time
dependent equation of motion of the radial component, when the
probe and the background branes are very close to each other.
We find out in particular exponential solution of the radial mode,
when the D1-brane approach the D5-branes in the pp-wave background.
For $p-q =2$, we find that there is a minimum distance beyond which
the D$q$-brane, which moves only radially, can't fall into the D$p$-branes.
One could possibly study some properties of the
branes by using other worldsheet techniques. \vskip .2cm \noindent
{\bf Acknowledgement:} \vskip .2cm
\noindent
KLP would like to
thank C. Becchi and C. Imbimbo for encouragement. The work of JK
was supported  in part by the Czech Ministry of Education under
Contract No. MSM 0021622409, by INFN, by the MIUR-COFIN contract
2003-023852, by the EU contracts MRTN-CT-2004-503369 and
MRTN-CT-2004-512194, by the INTAS contract 03-516346 and by the
NATO grant PST.CLG.978785.The work of RRN was supported by INFN.
The work KLP was supported partially by PRIN 2004 - "Studi
perturbativi e non perturbativi in teorie quantistiche dei campi
per le interazioni fondamentali".

\end{document}